\pdfoutput=1

\documentclass[11pt]{article}

\usepackage{acl}

\usepackage{times}
\usepackage{latexsym}
\usepackage{amsmath}

\usepackage{amssymb}
\usepackage{mathabx}
\usepackage{graphicx}
\usepackage{multirow}
\usepackage{multicol}
\usepackage{stfloats}
\usepackage{float}
\usepackage{caption}

\usepackage[T1]{fontenc}
\usepackage{xcolor}
\usepackage{url}
\usepackage{makecell}
\usepackage{pifont}
\usepackage{booktabs}
\usepackage{balance}

\usepackage{longtable}
\usepackage{caption}
\usepackage{multirow}
\usepackage{booktabs}

\usepackage[utf8]{inputenc}

\usepackage{microtype}

\usepackage{inconsolata}

\title{Improving Automated Distractor Generation for Math Multiple-choice Questions with Overgenerate-and-rank}

\author{
\bf Alexander Scarlatos$^{1*}$, Wanyong Feng$^{1*}$, Digory Smith$^2$, Simon Woodhead$^2$, Andrew Lan$^1$ \\
  University of Massachusetts Amherst$^1$, Eedi$^2$ \\
  \texttt{\{ajscarlatos, wanyongfeng, andrewlan\}@umass.edu} \\
  \texttt{\{digory.smith,simon.woodhead\}@eedi.co.uk} \\
  }

\begin{document}
\maketitle
\def\thefootnote{*}\footnotetext{These authors contributed equally to this work.}
\begin{abstract}
Multiple-choice questions (MCQs) are commonly used across all levels of math education since they can be deployed and graded at a large scale. A critical component of MCQs is the distractors, i.e., incorrect answers crafted to reflect student errors or misconceptions. Automatically generating them in math MCQs, e.g., with large language models, has been challenging. In this work, we propose a novel method to enhance the quality of generated distractors through overgenerate-and-rank, training a ranking model to predict how likely distractors are to be selected by real students. Experimental results on a real-world dataset and human evaluation with math teachers show that our ranking model increases alignment with human-authored distractors, although human-authored ones are still preferred over generated ones.
\end{abstract}

\section{Introduction and Related Work}
Multiple-choice questions (MCQs) are commonly used to assess student knowledge across all levels of education, including math, since they can accurately assess student knowledge while being easy to administer and grade at scale \cite{Nitko:96, Airasian:01, Kubiszyn:16}. An MCQ is comprised of a question stem and several answer options. The \textit{question stem} establishes the context and presents a problem for students to solve. Among the options, there exists a \textit{key}, which is the correct answer, and multiple \textit{distractors}, which are the incorrect answers specifically designed to reflect student errors or misconceptions. Although MCQs offer numerous advantages for assessing student knowledge, crafting high-quality distractors poses a significant challenge for teachers and educators. High-quality distractors should be sufficiently challenging so students do not quickly identify them as incorrect answers. Additionally, they should be designed to target specific errors or misconceptions, enticing students who make these errors or hold these misconceptions to choose them. This delicate balance makes the creation of such high-quality distractors a time and labor-intensive endeavor \cite{kelly2013traditional}.

Earlier works on automatic distractor generation for math MCQs use constraint logic programming \cite{tomas2013automatic} or manually crafted rules \cite{prakash2023q} to generate distractors. However, these methods are restricted to template-generated MCQs, which have limited applicability in a broader context. More recent work \cite{dave2021math} trains a neural network to solve math problems and sample incorrect answers as distractors. Not surprisingly, the generated distractors fail to capture student errors or misconceptions. The most recent works \cite{mcnichols2023exploring,feng2024exploring} explore this task using state-of-the-art large language models (LLMs), such as \texttt{ChatGPT}. The authors experiment with several different approaches, including few-shot in-context learning \cite{brown2020fewshot} and zero-shot chain-of-thought (CoT) prompting \cite{wei2022chain}, showing that LLMs can often generate distractors that are mathematically relevant to the MCQ. However, the overall alignment level with human-authored distractors that are thought to reflect student errors or misconceptions is not high. These works indicate a need to understand what errors or misconceptions are common among students and to use this information to improve the quality of generated distractors. 

\subsection{Contributions}
In this work, we propose a method to enhance the quality of generated distractors through overgenerate-and-rank.\footnote{Our code is publicly available at \url{https://github.com/umass-ml4ed/distractor-ranking-BEA}} Our novel ranking model evaluates the likelihood of each generated distractor being selected by real students. We train the ranking model via direct preference optimization (DPO) on pairwise preference pairs that compare the relative portion of students selecting one distractor over the other. This method can be augmented with existing distractor generation methods. 

We validate the effectiveness of this method through extensive experiments on a real-world math MCQ dataset. We find that the ranking model effectively selects distractors that students are more likely to select. In particular, it can improve the generated distractor quality of a fine-tuned \texttt{Mistral} model with 7B parameters to a similar level as that of \texttt{GPT-4} with CoT prompting, which is rumored to have up to 1T parameters. We also conduct human evaluations where we ask math teachers to rank and rate both LLM-generated and human-authored distractors. Results show that our ranking model's ranking and human ranking correlate with actual ranking defined by the portion of students selecting each distractor to a similar degree. Despite the improvements, LLM-generated distractors still do not match the quality of human-authored ones in reflecting student errors or misconceptions. 

\section{Methodology}
This section contains the details of the task definition and our over-generate-and-rank method.

\subsection{Task Definition}
We define an MCQ $Q$ as comprising a collection of elements, denoted as $Q = \{s, k, e_k, D, F, P\}$. Specifically, each MCQ includes a question stem $s$, a key $k$, an explanation for the key $e_k$, and a set of distractors $D$. Each distractor $d_i \in D$ is associated with a feedback message $f_i \in F$ provided to students upon selection. Moreover, for the key and every distractor, we have $p_i \in P$ as the portion of students who select this distractor (among all students who solve the MCQ).\footnote{All elements within \(Q\), except for \(P\), are formatted as strings, whereas \(P\) is formatted as numbers.} Similar to \cite{qiu2020automatic}, we define the distractor generation task as learning a function $g^\text{dis}$ that outputs a set of distractors $\hat{D}$ for an MCQ given the question stem, key, and its explanation, i.e., $g^\text{dis}(s, k, e_k) \rightarrow \hat{D}.$

\subsection{Pairwise Ranking}
In order to identify high-quality distractors for overgenerate-and-rank, we propose a ranking function that aligns with how likely distractors are to be selected by students. We define the ranking function as $r(s,k,e_k,d_i) \rightarrow \alpha_i \in \mathbb{R}$, where $\alpha_i$ is a relative score for distractor $d_i$. Our goal is to train $r$ such that higher scoring distractors are more likely to be selected by students, i.e., $\alpha_i > \alpha_j \rightarrow p_i > p_j$. We achieve this alignment by setting $\alpha_i$ to the log likelihood of $d_i$ under a ranking model $\mathcal{M}$, i.e., $\alpha_i = \log P_\mathcal{M}(d_i|s,k,e_k)$, where $\mathcal{M}$ is an autoregressive language model trained to generate distractors that are likely to be selected by students.

We initially fine-tune a model $\mathcal{M}_\text{SFT}$, where all distractors in the train set are used as labels for their corresponding questions. While $\mathcal{M}_\text{SFT}$ captures the likelihood of a distractor to appear in the data, it does not account for student behavior. We therefore train a model $\mathcal{M}_\text{DPO}$ via direct preference optimization (DPO) \cite{rafailov2024direct}, using all $|D| \choose 2$ pairs of distractors for each question where the distractor chosen more frequently by students is the preferred one in each pair. This aligns the model with student selections, and is motivated by recent successes of DPO in educational tasks \cite{scarlatos2024improving, kumar2024improving}.

We validate the effectiveness of this approach by calculating the \textit{ranking accuracy}, i.e., the percentage of distractor pairs in the test set where the predicted ranking agrees with actual student selection percentages. $\mathcal{M}_\text{SFT}$ and $\mathcal{M}_\text{DPO}$ result in ranking accuracies of $61.60\%$ and $65.84\%$, respectively; we use the latter in our experiments. While these numbers may appear low (random selection yields $50\%$), we note that the data is noisy and accuracy improves when there is a higher difference between selection percentages: $\mathcal{M}_\text{DPO}$ gets $74.31\%$ accuracy on pairs where the difference between selection percentages is more than $10\%$. Training details are in Supplemental Material Section \ref{app:ed}.

\subsection{Overgenerate-and-rank and baselines}

We instruct a base distractor generation model to overgenerate a set of $n$ distractors, $D'$, such that $n > |D|$.
Subsequently, we use our learned ranking model to score each candidate distractor $d_i \in D'$ and choose the $|D|$ distractors with the highest scores as our final set of generated distractors \cite{kumar2023improving}. In practice, we use $n=10$ and have $|D|=3$ (\textbf{Top-3}). We compare our method against two baseline ranking methods: First, we simply randomly select 3 distractors from $D'$ (\textbf{Rand-3}). Second, we instruct the base distractor generation model to directly generate exactly 3 distractors (\textbf{Only-3}).

\section{Experiments}
This section provides a comprehensive overview of our dataset, outlines the evaluation metrics and the experimental setup, and details the findings from experiments and human evaluation.
\subsection{Dataset}
We use a dataset that comprises 1.4K math MCQs sourced from Eedi's content repository\footnote{\url{https://eedi.com/home}}. These questions, all written in English, target students aged 10 to 13. Each MCQ includes a question stem, a key with an explanation justifying its correctness, and three distractors, each accompanied by a feedback message clarifying why it is incorrect. Additionally, each option is tagged with the percentage of students choosing that option, computed on an average of 4,000 student responses per question. We split the dataset into training and test sets at an 80:20 ratio. The training set is used to fine-tune the base distractor generation LLM (if necessary) and train the ranking model, while the test set is used for evaluation.

\subsection{Evaluation Metrics}
We adopt the alignment-based metrics previously introduced in \cite{mcnichols2023exploring} to assess the degree of alignment between LLM-generated distractors and human-authored ones. There are two binary metrics: \textbf{Partial} match, which checks if at least one LLM-generated distractor matches the human-authored ones\footnote{We use the exact string match criterion to align LLM-generated with ground-truth, human-authored distractors.}, and \textbf{Exact} match, which checks if all LLM-generated distractors match the human-authored ones. There is also one scalar metric: Proportional (\textbf{Prop.}) match, which calculates the proportion of LLM-generated distractors that match the human-authored ones. Additionally, to reflect the portion of students selecting each distractor, we introduce a new scalar metric: Weighted Proportional (\textbf{W. Prop.}) match (that also has range $[0,1]$), formally defined as
\begin{align*}
    h(D, \hat{D}) = \textstyle\sum_i I(\exists j \text{ s.t. } d_i = \hat{d_{j}}) \cdot p_i /\textstyle\sum_i p_i,
\end{align*}
where \(I\) is the indicator function. Intuitively, this metric re-weights each ``match'' in the Proportional metric such that a match on a distractor that more students select is weighed more heavily than one that less students select. We calculate the values for all metrics by averaging them across all MCQs in the test set and then scaling these values by a factor of 100 to convert them into percentages.

\subsection{Experimental Setup}
Following \cite{mcnichols2023exploring}, we use zero-shot chain-of-thought prompting (\textbf{CoT}) with \texttt{GPT-4} and fine-tuning (\textbf{FT}) with the open-source \texttt{Mistral}-7B model as our base distractor generation models. Since our goal is to evaluate the performance of the ranking model, we do not use their in-context learning method, ``kNN'', because in-context examples leak student selection information into the distractor generation model by showing example distractors that real students frequently select. Consistent with the best practices identified in their work, we represent each target MCQ by concatenating the question stem, the key, and its corresponding explanation. During the distractor generation process, the model must generate a feedback message before the actual distractor. Hyperparameters and model details are listed in the Supplementary Material Section~\ref{app:ed}. 

\subsection{Results and Discussion}
\begin{table}
\vspace{0.1cm}
\centering
\scalebox{.8}{
\begin{tabular}{{cccccc}}
\toprule
\multirow{2}{*}{\textbf{\textbf{Approach}}} &
  \multirow{2}{*}{} &
  \multirow{2}{*}{Partial} &
  \multirow{2}{*}{Exact} &
  \multirow{2}{*}{Prop.} &
  \multirow{2}{*}{W. Prop.} \\
\\ \midrule
                        & Top-3  &\textbf{67.87}  &2.53  &\textbf{32.25}  &\textbf{36.89}  \\
\multicolumn{1}{c}{CoT} & Rand-3 &47.29  &0.00  &18.29  &19.13  \\
                        & Only-3 &66.43  &\textbf{3.25}  &31.05  &35.03  \\
\midrule
                        & Top-3  &\textbf{67.15}  &1.44  &\textbf{30.20}  &\textbf{34.81}  \\
\multicolumn{1}{c}{FT}  & Rand-3 &35.38  &0.36  &14.20  &15.06  \\
                        & Only-3 &60.29  &\textbf{2.89}  &28.28  &31.75  \\
\bottomrule 
\end{tabular}
}
\caption{Results of distractor generation on alignment-based metrics. We see that overgenerate-and-rank (sometimes significantly) improves performance.}
\label{tab:result}
\end{table}

Table~\ref{tab:result} shows the performance of both base distractor generation models with different ranking methods across alignment-based metrics. The low Exact match values across methods indicate it is nearly impossible for the LLM to recover the exact 3 human-authored distractors. However, Top3 outperforms both Rand3 and Only3 on all other metrics, which suggests that the trained ranking model is effective at identifying which distractors are more likely selected by students. 
The gap on the Weighted Proportional metric is bigger than that on the Proportional metric for CoT and FT since the Weighted Proportional metric incorporates student distractor selection percentages, which is what the ranking model trains on.
This observation highlights the advantage of overgenerate-and-rank, suggesting that letting the base distractor generation model to generate a diverse set, casting a wide net, and then using the ranking model to select good ones is an effective approach.
Perhaps most importantly, we see that Top3 with FT performs similarly to Only3 with CoT. This observation shows that the ranking model can elevate the performance of a small, open-source LLM (\texttt{Mistral-7B}) and make it comparable to a much larger, proprietary LLM (\texttt{GPT-4}), which is a promising sign for the potential real-world deployment of automated distractor generation methods in a cost-controlled way. 

\subsection{Human Evaluation}

We conduct human evaluations where we recruit two math teachers with experience teaching grade-school-level math to evaluate distractors. We randomly select 20 MCQs whose Top-3 LLM-generated distractors are completely different from the human-authored ones from the test set. In the first evaluation task, we ask evaluators to rank the quality of human-authored distractors to examine the correlation between teacher judgment (\textbf{Human Rank}), the ranking model's ranking (\textbf{Model Rank}), and the actual student selection percentages (\textbf{GT Rank}). In the second evaluation task, we show evaluators 6 distractors for each MCQ, including 3 LLM-generated distractors and 3 human-authored distractors. We then ask them to rate the overall quality of each distractor to compare LLM-generated distractors (\textbf{Top-3 LLM}) with human-authored ones (\textbf{Human}), on a 5-point Likert scale, from 1 (least likely to be selected by students) to 5 (most likely). To mitigate potential bias from distractor ordering, the sequence of the distractors was randomized for each MCQ. 

\begin{table}[t!]
\centering
\scalebox{1.0}{
\begin{tabular}{@{}ccc@{}}
\toprule
\textbf{Comparison} & \textbf{Kendall's Tau} \\ \midrule
GT Rank vs. Human Rank & 0.27 \\
GT Rank vs. Model Rank & 0.30  \\
Human Rank vs. Model Rank & 0.14 \\ \bottomrule
\end{tabular}
}
\caption{Correlation between different rankings on human-authored distractors. Teachers and the ranking model correlate with actual student selection percentages to a similar degree.}
\label{tab:correlation}
\end{table}

\begin{table}[t!]
\centering
\scalebox{.88}{
\begin{tabular}{cccc}
\toprule
\multicolumn{2}{c}{\textbf{QWK}} & \multicolumn{2}{c}{\textbf{Average Ratings}} \\
        \cmidrule(lr){1-2} \cmidrule(lr){3-4}
Top-3 LLM & Human & Top-3 LLM & Human \\
\midrule
0.66        & 0.62          &  $2.67\pm 0.96$              & $3.26\pm 1.02^*$                 \\
\bottomrule
\end{tabular}
}
\caption{Inter-rater agreement and average ratings on LLM-generated and human-authored distractors. $*$ indicates statistical significance ($p<0.05$) under a t-test.}
\label{tab:human_eval}
\end{table}

\begin{table}[t!]
\centering
\scalebox{.95}{
\begin{tabular}{@{}ccc@{}}
\toprule
\textbf{Head-to-Head Rating Comparison} & \textbf{Percentage} \\ \midrule
$\text{Top-3 LLM} > \text{Human}$  & 22\% \\
$\text{Top-3 LLM} = \text{Human}$  & 16\% \\
$\text{Top-3 LLM} < \text{Human}$  & 62\% \\ \bottomrule
\end{tabular}
}
\caption{Head-to-head comparison between LLM-generated distractors and human-authored ones. Teachers prefer human-authored ones most of the time.}
\label{tab:head}

\end{table}

Table \ref{tab:correlation} shows Kendall's Tau correlation \cite{kendall} between the ground-truth ranking and the human/model ranking. We see that human and model rankings have a weak-to-moderate correlation with the ground-truth ranking. This observation reveals the difficulty of this task since even expert math teachers with years of teaching experience cannot fully anticipate real students' behavior. 
We also see that human ranking and model ranking have a weak correlation, likely due to humans and LLMs approaching the same problem from different angles; future work can consider a human-AI collaboration approach. 

Table \ref{tab:human_eval} shows the inter-rater agreement among math teachers, measured in quadratic weighted Kappa (QWK) \cite{qwk}, and their average ratings for both LLM-generated distractors and human-authored ones. We see that human-authored distractors are preferred with statistical significance, and the inter-rater agreement is moderate-to-substantial.
However, we note that since the 20 selected MCQs in our evaluation are the ones where none of the top-3 LLM-generated distractors match human-authored ones, this result may
downplay the effectiveness of LLMs because they must generate plausible distractors that are not already included in the human-authored ones.

We additionally compare the LLM-generated and human-authored distractors head-to-head, using average distractor rating across evaluators between each LLM-generated distractor and each human-authored distractor for each question (resulting in 9 comparisons per question). Table \ref{tab:head} shows the percentage of cases where LLM-generated distractors win, lose, or tie to human-authored ones. We see that even though human-authored distractors are preferred the majority of the time, there is a sizeable portion of LLM-generated distractors that are equal to or preferred over human-authored distractors. This result implies that LLMs can generate some high-quality distractors that can be used to enhance the quality of human-authored ones.

\section{Conclusions and Future Work}

In this paper, we propose an overgenerate-and-rank method for generating distractors for math MCQs. We train a ranking model to predict which distractors students would select more often, and this ranking model can be applied to any existing distractor generation method. We experimentally validate its performance on a real-world dataset and test its limitations through human evaluation. 

Avenues for future work include but are not limited to further improving the ranking model through a student-specific distractor selection prediction objective that considers their knowledge state \cite{liu2022open}, developing a human-in-the-loop approach for distractor selection percentage prediction, and using the same approach for feedback generation \cite{scarlatos2024improving}. Finally, extending our work from multiple-choice questions to open-ended questions is important, since open-ended student responses contain much more detailed information on their errors \cite{zhang2021math,zhang2022automatic,mcnichols2023algebra}. 

\section*{Limitations}
First, due to limited resources, we only performed human evaluation on the human-authored distractors and the Top-3 LLM-generated distractors. However, this does not allow us to determine if our overgenerate-and-rank approach is better than generation baselines from a human evaluation perspective. We also acknowledge that our human evaluation sample size is small, and should ideally be increased for future studies. Second, while we have evidence that our method enhances the quality of LLM-generated distractors, a notable difference remains between the quality of LLM-generated distractors and human-authored ones. To make LLM-generated distractors viable for deployment in real educational settings, it is necessary to further investigate how to improve their overall quality. Third, our first human evaluation result shows that even experienced math teachers cannot anticipate real student behavior accurately. A more precise evaluation for LLM-generated distractors would involve deploying them in actual tests and observing student behavior. However, this process can be significantly complicated and time-consuming, and should only be performed when there is reasonable evidence that generated distractors might be of similar quality to human-authored ones.

\section*{Ethical Considerations}
Our work uses the overgenerate-and-rank method to improve the quality of LLM-generated distractors. We believe that our work could potentially reduce the time educators and teachers spend creating math MCQs, enabling them to focus more on teaching and engaging with students. However, we acknowledge that potential biases within LLMs may exist, which could cause the LLM-generated distractors to contain incorrect or potentially harmful information. Therefore, we strongly recommend that educators and teachers review the quality of LLM-generated distractors thoroughly before deploying them in actual tests for students.

\section{Acknowledgement}
We thank Schmidt Futures and the NSF (under grants IIS-2118706 and IIS-2237676) for supporting this work. We also thank Craig Barton, Claire Willis, and Rachel Kidson for their insights on the distractor ranking task. 

\bibliography{anthology, custom}

\clearpage
\newpage 

\noindent {\Large \textbf{Supplementary Material}}

\appendix

\section{Distractor Generation Examples}
\begin{table} [h!]
    \centering
    \scalebox{0.85}{
    \begin{tabular}{p{8.5cm}}
    \toprule
    \textbf{Question Stem} \\
    fifty five thousand subtract twenty three thousand equals \\
    \midrule
    \textbf{Key}\\ 32,000\\
    \midrule
    \textbf{Human-authored Distractors}\\ 22,000 $\quad$ 23,000 $\quad$ 3,200 $\quad$\\
    \midrule
    \textbf{Only-3 LLM-generated Distractors}\\ 52,000 $\quad$ -32,000 $\quad$ 78,000 $\quad$\\
    \midrule
    \textbf{Top-3 LLM-generated Distractors}\\ 33,000 $\quad$ -32,000 $\quad$ 30,000 $\quad$\\
    \bottomrule
    \end{tabular}
    }
    \label{table:distractor-example-fraction-simplify}
    \caption{Representative question with distractors from humans, GPT-4 generating only 3, and GPT-4 after selecting the top 3 with our ranking model.}
\end{table}

\section{Experimental Details}
\label{app:ed}

We take several measures to ensure that generated distractors are distinct and different from the key. For CoT, we prompt \texttt{GPT-4} to generate 15 distractors and eliminate duplicates and those identical to the key. For the rest of MCQs lacking 10 distinct distractors, we prompt \texttt{GPT-4} again to generate 15 new distractors, instructing it to avoid producing previously generated distractors by including them in the prompt. We supplement the existing distractors with the newly generated distractors, ensuring the total number of distinct distractors reaches 10. For the MCQs that still lack 10 distinct distractors (which are few), we add the word "placeholder" as distractors. We use greedy decoding for all previous steps.
When overgenerating distractors with our fine-tuned model, we generate 3 distractors 5 times using nucleus sampling for each MCQ, setting $\text{temperature}=1$ and $\text{top\_p}=0.9$. If we do not get 10 unique distractors, we generate 5 more times with $\text{top\_p}=1.0$ to ensure greater diversity. When generating only 3 distractors, we use beam search with $\text{num\_beams}=5$. If we do not get 3 unique distractors, we then generate with nucleus sampling twice with $\text{top\_p}=0.9$ and take the first 3 unique distractors.

For the fine-tuned distractor generation model and the pairwise ranking model, we use the \texttt{mistralai/Mistral-7B-v0.1} model from HuggingFace \cite{huggingface} and load the model with 8-bit quantization \cite{int8}. We train LoRA adapters \cite{lora} on the \texttt{q\_proj}, \texttt{v\_proj}, \texttt{k\_proj}, and \texttt{o\_proj} matrices, setting $r=32$, $\alpha=16$, $\text{dropout}=0.05$. We train using the AdamW optimizer with a virtual batch size of 64 using gradient accumulation and do early stopping on a random $20\%$ subset of the train set. For the distractor generation model we use a learning rate of 5e-5 and train for 15 epochs, and for the pairwise ranking model we use a learning rate of 3e-5 and train for 5 epochs. For DPO training on the pairwise ranking model, we set $\beta=0.5$ and use $\mathcal{M}_\text{SFT}$ as the reference model. We train all models on a single NVIDIA RTX A6000 GPU.

\section{Human Evaluation Details}
\label{app:he}
In this work, we obtained approval from the ethics review board for human evaluation. We show the evaluation instructions to human evaluators in Table ~\ref{tab:ins}. We do not provide any compensation for human evaluators because their participation is entirely voluntary and we appreciate their contribution to this work.

\clearpage
\onecolumn
\section{Prompt Format}
We provide the prompts for CoT, FT, and pairwise ranking model below. We use $<>$ to indicate that a variable is filled in dynamically.
\begin{longtable}{p{4.5cm}p{10.5cm}}
    \toprule
    \textbf{Prompt} & You are provided with a math question, correct answer, and the explanation of correct answer. Your task is to use the following template to create 15 unique incorrect answers (distractors) to be used as multiple-choice options for a middle school math multiple-choice question. Before generating each distractor, include a concise explanation to clarify for students why that is not the correct answer. Make sure each distractor is clearly different from the correct answer and distinct from each other, this is very important!\newline 
    [Template]\newline 
    Distractor1 Feedback:\newline
    Distractor1:\newline
    Distractor2 Feedback:\newline
    Distractor2:\newline
    Distractor3 Feedback:\newline
    Distractor3:\newline
    Distractor4 Feedback:\newline
    Distractor4:\newline
    Distractor5 Feedback:\newline
    Distractor5:\newline
    Distractor6 Feedback:\newline
    Distractor6:\newline
    Distractor7 Feedback:\newline
    Distractor7:\newline
    Distractor8 Feedback:\newline
    Distractor8:\newline
    Distractor9 Feedback:\newline
    Distractor9:\newline
    Distractor10 Feedback:\newline
    Distractor10:\newline
    Distractor11 Feedback:\newline
    Distractor11:\newline
    Distractor12 Feedback:\newline
    Distractor12:\newline
    Distractor13 Feedback:\newline
    Distractor13:\newline
    Distractor14 Feedback:\newline
    Distractor14:\newline
    Distractor15 Feedback:\newline
    Distractor15:\newline    
    Question: <question>\newline
    Explanation: <explanation>\newline
    Answer: <answer>\\
    \bottomrule
    \caption{Prompt for chain-of-thought distractor generation with GPT-4.}
    \label{tab:prompt-CoT} \\
\end{longtable}

\clearpage
\onecolumn

\begin{longtable}{p{4.5cm}p{10.5cm}}
    \toprule
    \textbf{Prompt} & You are provided with a math question, correct answer, and the explanation of correct answer. Your task is to generate 3 unique incorrect answers (distractors) to be used as multiple-choice options for a middle school math multiple-choice question. Before generating each distractor, include a concise explanation for students to clarify why that is not the correct answer. Ensure each distractor is different from the correct answer and distinct from the others; this is very important!\newline    
    Question: <question>\newline
    Explanation: <explanation>\newline
    Answer: <answer>\\
    \bottomrule
    \caption{Prompt for fine-tuning with Mistral.}
    \label{tab:prompt-FT} \\
\end{longtable}

\begin{longtable}{p{4.5cm}p{10.5cm}}
    \toprule
    \textbf{Prompt} & A teacher assigns the following math question to a class of middle school students. \newline
    \newline
    Question: <question> \newline
    Solution: <solution> \newline
    Correct answer: <correct answer> \newline
    Generate a distractor for this question that targets some student misconception. \newline
    \newline
    Distractor: <distractor>\\
    \bottomrule
    \caption{Prompt for pairwise ranking model.}
    \label{tab:prompt-pairwise} \\
\end{longtable}

\clearpage
\onecolumn
\section{Human Evaluation Instructions}
\begin{longtable}{p{15.5cm}p{10.5cm}}
    You are provided with two tasks \\
    The first task (rank) consists of 20 items, each containing a question stem and three distractors. For each item, you are asked to rank the three distractors based on the assessment of how often they will be selected by real students, from most frequent to least frequent. The items for this task can be accessed in the rank.csv file. \\
    Example: \\
    Question: How do you write 4.6 as a percentage? \\
    Distractor 1 (id = 1): 46\% \\
    Distractor 2 (id = 2): 0.046\% \\
    Distractor 3 (id = 3): 4.6\% \\
    Best distractor id: 1 \\
    Second best distractor id: 3 \\
    Third best distractor id: 2 \\
    \\
    The second task (rate) also consists of 20 items, each containing a question stem and six distractors. For each item, you are asked to rate the likelihood of each distractor being selected by students on a 5-point scale independently: 5 - most likely, 4 - likely, 3 - average, 2 - not likely, and 1 - least likely. The items for this task can be accessed in the rate.csv file \\
    Example: \\
    Question: How do you write 4.6 as a percentage? \\
    Distractor: 46\% \\
    Rating: 4 \\
    \caption{Instructions given to human evaluators for evaluating distractors.}
    \label{tab:ins}
\end{longtable}

\end{document}